\documentclass[prd,nobibnotes,nofootinbib,12pt]{revtex4-1}
\pdfoutput=1

\usepackage{amssymb}
\usepackage{amsmath}
\usepackage{graphicx}
\usepackage{verbatim}
\usepackage{amsfonts}
\usepackage{lipsum}

\allowdisplaybreaks

\usepackage{color}

\newcommand{\be}{\begin{equation}}
\newcommand{\ee}{\end{equation}}

\begin{document}

\vspace{-14mm}
\begin{flushright}
        {AJB-18-3}
\end{flushright}

\vspace{8mm}

\title{Analytic inclusion of the scale dependence of the anomalous dimension matrix in Standard Model Effective Theory}

\author{Andrzej J. Buras}
\email{andrzej.buras@tum.de}
\affiliation{TUM Institute for Advanced Study, Lichtenbergstr.~2a, D-85748 Garching, Germany}
\affiliation{Physik Department, TU M\"unchen, James-Franck-Stra{\ss}e, D-85748 Garching, Germany}
\author{Martin Jung}
\email{martin.jung@tum.de}
\affiliation{Excellence Cluster Universe, Technische Universit\"at M\"unchen, Boltzmannstr. 2, D-85748 Garching, Germany}
\vspace*{1cm}
\begin{abstract}
The renormalization group equations (RGEs) in Standard Model effective theory are usually either solved analytically,
neglecting the scale dependence of gauge and Yukawa couplings, or numerically without such approximations. We present
analytic solutions of RGEs that take into account the dominant scale dependence of the anomalous-dimension matrix due to
the running of the QCD coupling $\alpha_s$ and of the top-Yukawa coupling.
We consider first the 
case for which a given operator is generated directly through mixing with the {\it parent} operator whose Wilson coefficient is
non-vanishing at the new physics scale. Subsequently we consider the case of two-step running, in which two operators do not mix
directly, but only via a third {\it mediator operator}. We generalize  these solutions to an arbitrary number of operators and show how
even in this case analytic solutions can be obtained. 
\end{abstract}

\maketitle

%%%%%%%%%%%%%%%%%%%%%%%%%%%%%%%%%%%%%%%%%%%%%%%%%%%%%%%%%%%%%%%%%%%%%%%%%%%%%%%%%%%%%%%%%%%%%%%%%%%%%%%%%%%%%%%%%%%%%%%%%%%%%%%%%%%%
%%%%%%%%%%%%%%%%%%%%%%%%%%%%%%%%%%%%%%%%%%%%%%%%%%%%%%%%%%%%%%%%%%%%%%%%%%%%%%%%%%%%%%%%%%%%%%%%%%%%%%%%%%%%%%%%%%%%%%%%%%%%%%%%%%%%

\section{Introduction}

The absence of any signal in direct searches for particles beyond the Standard Model (SM) at the LHC renders a scenario likely in
which the scale $\Lambda$ of such potential particles is much larger than the electroweak scale. The resulting hierarchy can
be used together with the assumption of linearly realized electroweak symmetry breaking to formulate the so-called Standard Model
Effective Theory (SMEFT), provided no undiscovered weakly-coupled {\em light} particles exist, like axions or sterile neutrinos. 
This framework incorporates the full SM gauge symmetry which is unbroken at the high scale $\Lambda$ and allows to investigate its
implications on observables measured at scales significantly below $\Lambda$. Technically this is realized via an operator product
expansion, where all the SM degrees of freedom are kept as dynamical degrees of freedom and only the NP is integrated out, yielding
dimension-six operators built out of SM fields that are invariant under the full SM gauge symmetry.
This type of effective theory is usually called  SMEFT, because it reduces at low energies to the SM. Yet, the presence of
dimension-six operators whose Wilson coefficients can in principle be arbitrary can introduce very significant modifications to
the SM phenomenology.

The effective Lagrangian of SMEFT
at dimension six is given as follows:
\begin{equation}\label{BASICSMEFT}
\mathcal L^{(6)} =  \sum_{k} C_k^{(6)} Q_k^{(6)}\,,
\end{equation}
with the contributing operators classified in full generality in \cite{Buchmuller:1985jz,Grzadkowski:2010es}, the latter
article for the first time providing an irreducible basis which is now standard.
The corresponding renormalization group analysis at leading order of all these operators has been carried out in
\cite{Jenkins:2013zja,Jenkins:2013wua,Alonso:2013hga}.\footnote{See \texttt{http://einstein.ucsd.edu/smeft/} for errata.}
The renormalization group equations (RGEs) in SMEFT derived there often involve many operators mixing with each other. The
corresponding  anomalous dimension matrix (ADM) depends not only on the three gauge couplings of the SM, but also on fermion Yukawa
couplings, in particular the top-quark Yukawa coupling, and quartic Higgs couplings.

Given this complicated structure, solutions to RGEs in SMEFT are typically obtained in one of two ways: either analytically,
assuming the ADM to be constant and furthermore commonly considering only the first leading logarithm, or numerically as
in \cite{Celis:2017hod,Aebischer:2018bkb,Workgroup:2017myk}, taking the scale-dependence of the ADM into account, but making
further analytical insight difficult. Although the former procedure can be useful in showing the overall importance of renormalization
group effects, see for instance
\cite{Hisano:2012cc,deBlas:2015aea,Cirigliano:2016njn,Cirigliano:2016nyn,Feruglio:2017rjo,Bobeth:2016llm,Bobeth:2017xry,Fuyuto:2017xup},
we prefer to avoid the assumption of a scale-independent ADM, since the anomalous dimensions involved in fact typically show a pronounced
scale dependence. However, analytical solutions are very useful, since they are easy to use and facilitate applying constraints at
different scales. Moreover, they give some insight into the pattern of the dominant dynamical effects.
The main goal of our paper is therefore to provide analytic solutions for the  relevant RGEs, \emph{including} the
phenomenlogically relevant effects from the scale dependence of the ADM, specifically the running of the \emph{strong} and
relevant \emph{Yukawa} couplings. In order to achieve this and to resum the leading logarithms to all orders in perturbation theory, we
exploit a distinct hierarchy in the different running effects:
\begin{enumerate}
\item
The running of the strong and top-Yukawa couplings themselves is dominated by SM physics; while effects from NP are
possible, they can be safely assumed to be at most of the order of the neglected NLO contributions.
\item For operators present at the scale $\Lambda$, the RG effects due to the latter two couplings are dominant, since they are the
only $\mathcal O(1)$ couplings. 
\item The scale-dependence of $g_1$ and $g_2$ is much weaker than that of $g_3$ (or equivalently $\alpha_s$) and $y_t$; we hence
consider $g_{1,2}$ fixed to their values at the electroweak scale.
\end{enumerate}

We therefore solve the RGEs in a step-wise fashion: we first discuss the scale dependence of the strong and top Yukawa couplings.
We use these results to solve the self-mixing of the \emph{parent} operators generated at the NP scale $\Lambda$. Finally, taking the
latter solution into account, we consider the running of the \emph{child} operators generated via weak mixing, including their
self-mixing.

Solving the RGEs including the running of both $\alpha_s$ and $y_t$ is not trivial, even with this hierarchy of different
effects; while we discuss the exact inclusion of both couplings, we achieve a much simpler form due to the crucial observation that
\be\label{eq::approx}
\frac{y_t^2(\mu)}{\alpha_s(\mu)}\approx {\rm const.}
\ee
The weak $\mu$-dependence of the right-hand side at leading order in $\alpha_s$ is discussed below and shown to have a negligible
effect, such that both effects can be included using a simple, very accurate approximation. Higher order corrections are unimportant at
scales $\mu \geq \mu_{\rm EW}$ considered by us. They would become more important at $\mu\ll \mu_{\rm EW}$, but $y_t(\mu)$
does not run at these scales as the top quark is integrated out at $\mu\approx\mu_{\rm EW}$. Nevertheless, the relation still
holds approximately for the remaining Yukawa couplings.

Our paper is organized as follows: In Section~\ref{sec:2} we recall the relevant formulae for RGEs and specify our assumptions.  In
Section~\ref{sec:3} we recall for pedagogical reasons the solutions of RGEs when the ADM is assumed to be scale independent. Here we
consider also the interesting case mentioned in the abstract when the coefficient of the first leading logarithm vanishes so that the
solution involves the square of $\ln(\Lambda/\mu)$.
In Section~\ref{sec::RGEsolution}, the main section of our paper, we derive the analytic solutions of RGEs under the assumptions
listed above. We demonstrate our method first for the case of a single non-vanishing Wilson coefficient at $\mu=\Lambda$ and
generalise it systematically to an arbitrary number of operators. The examples given there demonstrate the precision of our
analytical formulae. Our brief numerical analysis in Section~\ref{sec:5} demonstrates the application of the developed framework
very explicitly, in order to facilitate its application.
In Section~\ref{sec:6} we summarize our results by presenting Table~\ref{tab::summary}, which guides the reader to the main results of
our paper and should simplify their usage.

\section{Basic Setup}\label{sec:2}
The one-loop RGE can generically be written as 
\begin{equation}\label{RG}
\dot{\mathbf C}\equiv 16\pi^2\frac{d \mathbf{C}}{d\ln\mu} = \hat\gamma(\mu)\, \mathbf C\,,
\end{equation}
where $\mathbf C=(C_1,C_2,C_3,...)^T$  contains the Wilson coefficients of contributing operators and $\hat\gamma$ is a general,
scale-dependent anomalous dimension matrix that depends on gauge and Yukawa couplings.
Note that we do not use the traditional notation with the transpose of the anomalous dimension matrix in order to agree with
\cite{Jenkins:2013zja,Jenkins:2013wua,Alonso:2013hga}, where anomalous dimensions are defined for Wilson coefficients and not
operators. In our examples we will mostly use the conventions of these papers specifying exceptions.

We first consider a typical scenario in which NP generates a subset of parent operators ${\cal O}_i$ of the SMEFT at some high
scale $\Lambda\gg \mu_{EW}$. The RG-running of this subset down to $\mu_{EW}$, described by the RGEs in Eq.~\eqref{RG}, has the
following implications:
\begin{itemize}
\item Creation of non-vanishing coefficients of the child operators ${\cal O}_k$. In order to distinguish these operators from parent
operators ${\cal O}_i$, we use the indices $(k,l)$ for the former, while $(i,j)$ are used for the latter.
\item Mixing between the parent operators ${\cal O}_i$ (including their self-mixing), modifying their coefficients.
\item Mixing between the child operators ${\cal O}_k$ (including their self-mixing), modifying their coefficients.
\item Mixing of the child operators back into the parent operators.
\end{itemize}
Considering electroweak, strong and Yukawa interactions, the RG effects show additional hierarchies:
\begin{itemize}
\item \emph{Strong mixing} due to the QCD and (top) Yukawa couplings can generate large effects, and exhibits a strong scale
dependence. Note that these two contributions are qualitatively different: while Yukawa interactions can provide chiral flips, this is
not possible for QCD. We define parent operators to include all operators with $C_i(\Lambda)\neq 0$, together with those generated
from the ${\cal O}_i$ via strong mixing.
\item The child operators ${\cal O}_k$ are hence by definition generated via \emph{weak mixing}. The corresponding anomalous
dimensions include electroweak gauge couplings and/or Yukawa couplings from lighter fermions. As such, they fulfill generically
$\gamma_{\rm weak}/\gamma_{\rm strong}\ll 1$. Analogously to parent operators, the class of child operators includes also those
not directly generated via weak mixing, but via strong mixing from operators generated via weak mixing. As already stated in the
introduction, the scale dependence of $g_{1,2}$ will be neglected.
\item Since by definition $C_k(\Lambda)\equiv 0$, the mixing of child into parent operators is of higher order, even without
additional hierarchies. Given the definitions above, it is actually at least suppressed as $\gamma_{\rm weak}^2/\gamma_{\rm strong}$
relative to strong-mixing contributions, and hence negligible.
\end{itemize}
Schematically, we obtain the following block form for the ADM, writing $P$ and $C$ for parent and child operators, respectively:
\begin{equation}\label{eq::block}
\hat\gamma = \left(\begin{array}{cc}\hat\gamma_{\rm strong}^{P} & 0\\\hat\gamma_{\rm weak}^{P\to C} & \hat\gamma_{\rm strong}^{\rm C}
\end{array}\right)\,,\quad{\rm with}\quad \mathbf{C}(\mu)=\left(\begin{array}{c}\mathbf{C}_{\rm P}(\mu)\\\mathbf{C}_{\rm C}(\mu)\end{array}\right)\quad{\rm and}\quad
\mathbf{C}(\Lambda)=\left(\begin{array}{c}\mathbf{C}_{\rm P}(\Lambda)\\0\end{array}\right)\,.
\end{equation}
Our goal is to include all phenomenologically relevant effects in this setup.

Formally, the solution of Eq.~\eqref{RG} can be written as
\begin{equation}\label{RG1}
\mathbf C(\mu) = \exp\left[\int_{\ln\Lambda}^{\ln\mu}\frac{\hat\gamma(\tilde\mu)}{16\pi^2}\,d\ln\tilde\mu\right]\mathbf C(\Lambda)\,,
\end{equation}
and the result for $\mathbf C(\mu)$ can in principle be obtained by numerically performing the integral in the exponential. In
this manner all the effects discussed above can be taken into account. Moreover, as the result is written in terms of an exponential,
all leading logarithms are summed up to all orders of perturbation theory. It should be remarked that in the presence of two-loop ADMs
(\ref{RG1}) should be generalized to include $T_g$ ordering that takes into account that one-loop and two-loop ADMs generally do not
commute with each other.  However, in this paper we will confine our discussion to one-loop ADMs.

\section{Scale-independent ADM}\label{sec:3}

We begin with  the simplest scenario in which $\hat\gamma$ is scale independent to find
\begin{equation}\label{RG2}
\mathbf C(\mu) = \exp\left[-\frac{\hat\gamma}{16\pi^2}\ln\!\left(\frac{\Lambda}{\mu}\right)\right]\mathbf C(\Lambda)\,.
\end{equation}
The leading logarithms are summed to all orders, but assuming $\hat\gamma$ to be scale independent is an approximation that will
be remedied in the next section.

\subsection{One step running}

If the argument of the exponential in Eq.~\eqref{RG2} is sufficiently below unity, we can expand it to obtain
\begin{equation}\label{RG3}
\mathbf C(\mu) = \left[\hat 1-\frac{\hat\gamma}{16\pi^2}\ln\!\left(\frac{\Lambda}{\mu}\right)\right]\mathbf C(\Lambda)\,,
\end{equation}
with $\hat 1$ denoting the unit matrix. This result is often encountered in the literature. 
The effects included in this rough solution can be made explicit by simply performing the multiplication of $\hat\gamma$ and $\mathbf
C(\Lambda)$, keeping our conventions for the indices $(i,j,k)$ in mind:
one  finds ($\mu < \Lambda$)
\be\label{simplest}
C_k(\mu)=-\frac{\hat\gamma_{ki}}{16\pi^2}\ln\!\left(\frac{\Lambda}{\mu}\right)
 C_i(\Lambda)\,, \qquad C_j(\mu)=\left[\delta_{ji}-\frac{\hat\gamma_{ji}}{16\pi^2}\ln\!\left(\frac{\Lambda}{\mu}\right)\right]
 C_i(\Lambda)\,.
\ee
We observe the following:
\begin{itemize}
\item The first result above describes the generation of child operators from parent operators due to weak mixing.
\item The second one describes the evolution of parent operators due to strong mixing, including self mixing due to $\hat\gamma_{ii}$.
\item While the self mixing affects the values of the parent coefficients $C_i(\mu)$, it does not have impact on child coefficients
$C_k(\mu)$ at this order.
\item Similarly, the strong mixing among the child coefficients has no impact on other child coefficients at this order, even
if $\hat\gamma_{kl}\not=0$.
\end{itemize}

The latter two points can be improved by expanding Eq.~\eqref{RG2} to higher orders (or using the full solution),  
although effects not taken into account in Eq.~\eqref{simplest} and listed above will generally be subleading. 
Generally higher powers of $\hat\gamma_{\rm weak}$ will be neglected in this expansion; however, such contributions can be
relevant when the contributions at first order are absent or heavily suppressed, \emph{e.g.} by several Yukawa couplings of light
fermions. We will discuss  this case now, still without the inclusion of the scale-dependence in $\hat\gamma$.

\subsection{Two-step running}

The first formula in Eq.~\eqref{simplest}  tells us that for a given element ${\hat\gamma_{ki}}=0$ no mixing occurs between
the operators $\mathcal{O}_k$ and $\mathcal{O}_i$ at one-loop level, and consquently $C_k(\mu)=0$ in ordinary  perturbation
theory. On the one hand, such mixing could take place at two-loop level, in which case the contribution would be of the order
$\gamma_{\rm weak}^2\ln(\Lambda/\mu)$. However, it turns out that the renormalization group improved solution \eqref{RG2} can
generate non-vanishing $C_k(\mu)$ through the so-called {\it two-step running}, even if ${\hat\gamma_{ki}}=0$ at one-loop. Such
contributions receive an additional enhancement by $\ln(\Lambda/\mu)$ compared to the two-loop matching contribution, which can
render them dominant for high scales $\Lambda$. We consider only these enhanced contributions in the following. This mechanism was
primarily discussed in the context of electric dipole moments \cite{Hisano:2012cc,Cirigliano:2016njn,Cirigliano:2016nyn}, where a
situation occurs when a given operator of interest does not mix directly with a second operator that enters an experimental observable,
but does so via a third \emph{mediator operator}. In spite of the presentations in
\cite{Hisano:2012cc,Cirigliano:2016njn,Cirigliano:2016nyn} we describe this case again in an attempt to exhibit the resulting
structure more clearly.

To be specific, we consider the following situation: the coefficient of an operator $\mathcal O_i$ is the only one with a non-zero
value at some high scale $\Lambda$. At some much lower scale $\mu$ an observable is determined by the value of a coefficient of an
operator $\mathcal O_k$, which does not mix directly with $\mathcal O_i$, but $\mathcal O_i$ mixes into a third operator $\mathcal O_m$
and $\mathcal O_m$ mixes into $\mathcal O_k$. The one-loop RGE and its solution in the approximation of scale-independent $\hat\gamma$
are again given by Eqs.~\eqref{RG} and~\eqref{RG2}, respectively, where additionally
$\mathbf C=(C_i,C_m,C_k)^T$ and 
\begin{equation}\mathbf C(\Lambda)=(C_i(\Lambda),0,0)^T\,,\quad \hat\gamma_{ki}=0\,,\,\hat\gamma_{mi}\not=0\quad \mbox{and}\quad
\hat\gamma_{km}\not=0\,.\label{eq::initial}
\end{equation}
Expanding the exponential, we find that a non-vanishing result for $C_k(\mu)$ is obtained first at second order which introduces a
factor $1/2$ and a logarithm squared. One can also check that as long as Eqs.~\eqref{eq::initial} are satisfied, the result for
$C_k(\mu) $ is independent of other entries in $\hat\gamma$. The latter enter first at third order and can be neglected. 
We thus find the leading contribution
\begin{equation}\label{eq::C3}
C_k(\mu)=\frac{1}{2}\hat\gamma_{km}\,\hat\gamma_{mi}\left[\frac{1}{16\pi^2}\ln\left(\frac{\Lambda}{\mu}\right)\right]^2C_i(\Lambda)\,.
\end{equation}
This result is trivially extended to block-diagonal forms of $\hat\gamma$. 
Needless to say that  Eq.~\eqref{RG2} allows for the generalization of this result to an arbitrary number of operators and to
contributions via arbitrary levels of mediator operators, which, however, will typically give negligible contributions.

\section{Analytical solution of RGEs \label{sec::RGEsolution}}

\subsection{The case of a single Wilson coefficient}\label{simple}
\begin{figure}
\includegraphics[width=9cm]{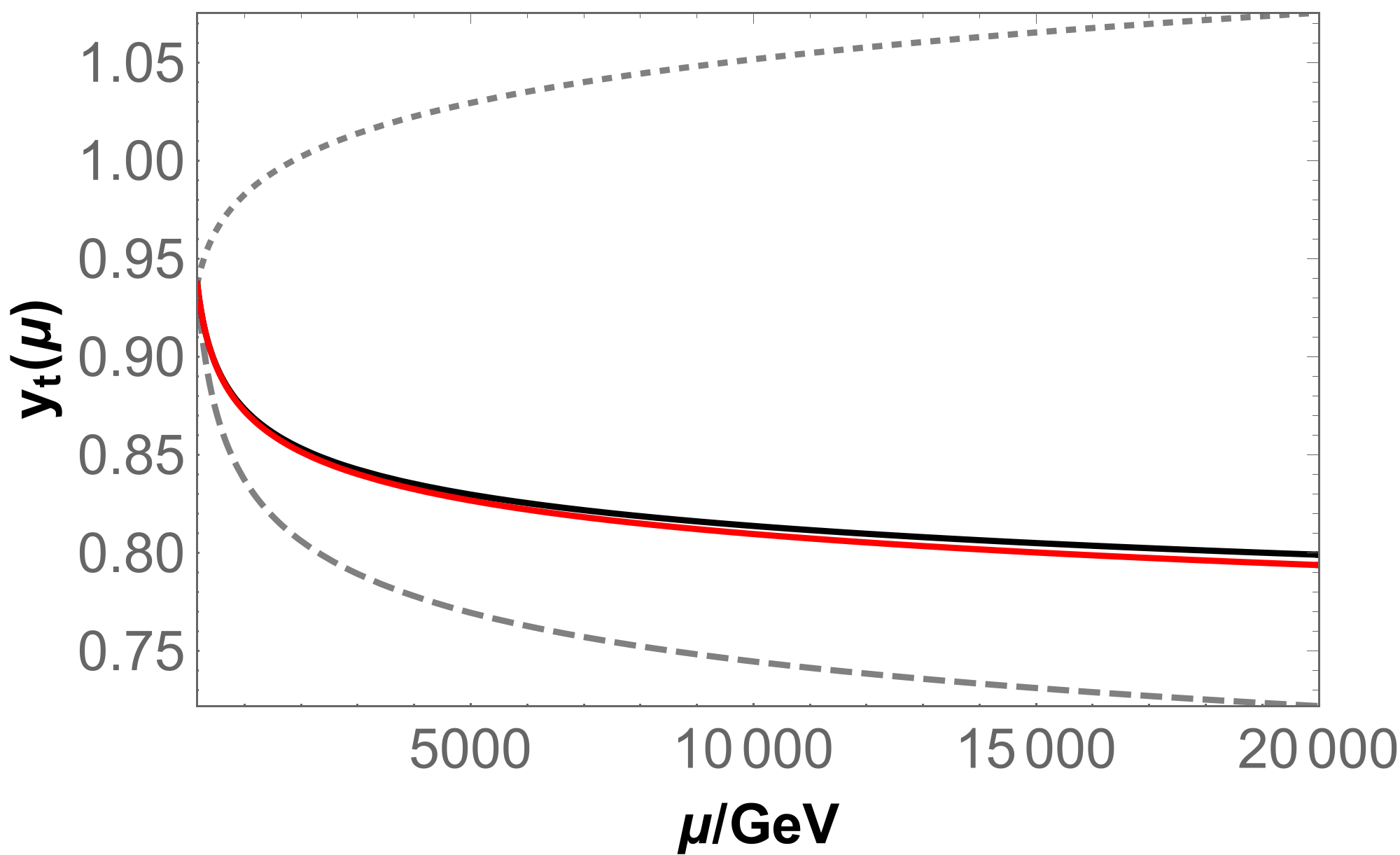}
\caption{\label{fig::ytrunning} The running of the top Yukawa coupling in different approximations: grey-dashed: only $\alpha_s$
running; grey-dotted: only $y_t^3$ running; black: exact solution \cite{Balzereit:1998id}; red: our approximation, see
Eqs.~\eqref{eq::yqrunning},\eqref{eq::basicrunning2}.}
\end{figure}

We will next take into account the scale dependence of $\hat\gamma$ resulting from $\alpha_s(\mu)$ and $y_q(\mu)$. 
For pedagogical reasons we start with the derivations for a single coefficient ${\cal C}_i$ with the corresponding anomalous
dimension $\gamma_i(\mu)$ to be explicitly given below. The relevant RG equation remains Eq.~\eqref{RG} and its formal
solution Eq.~\eqref{RG1}, both written for a single coefficient.
To perform the integration in Eq.~\eqref{RG1}, we first reiterate that the only two numerically important
contributions in self mixing are the ones from the strong and the top-Yukawa couplings that can both appear in $\gamma_i(\mu)$.
Regarding the scale dependence of the strong coupling, we use the standard leading-logarithmic solution
\begin{equation}\label{eq::basicrunning1}
\alpha_s(\mu) = \frac{\alpha_s(\mu_0)}{1+\alpha_s(\mu_0)\frac{\beta_0}{2\pi}\ln\frac{\mu}{\mu_0}}\,,
\end{equation}
with some reference scale $\mu_0$.
As far as the top-Yukawa coupling is concerned, we note that its QCD evolution satisfies for $\mu > \mu_\text{EW}$ to an
excellent approximation
\be\label{MAIN}
\frac{y_t^2(\mu)}{\alpha_s(\mu)} \approx {\rm constant}.
\ee
For the top Yukawa, we also take a non-linear term $\sim y_t^3$ \cite{Cheng:1973nv,Machacek:1983fi} approximately into account. 
Assuming Eq.~\eqref{MAIN} to hold for the full solution, we obtain\footnote{Note that the scale $\mu_0$ does not have to be equal
to the one in $\alpha_s(\mu)$.}
\be\label{eq::yqrunning}
y_q(\mu) = y_q(\mu_0)\left[\eta(\mu,\mu_0)\right]^{\epsilon_{y_q}}\,,
\ee
%.
where we introduced
\be\label{eq::basicrunning2}
\eta(\mu,\mu_0)\equiv \frac{\alpha_s(\mu)}{\alpha_s(\mu_0)},\qquad \epsilon_{y_q} \equiv
\frac{1}{8\pi\beta_0}\,\left[4\pi\gamma_m^{(0)}-\frac{9}{2} \frac{y^2_q(\mu_0)}{\alpha_s(\mu_0)}\right]\,,
\ee
with $\gamma_m^{(0)}= 4 C_F=8$.
The expression for $\epsilon_{y_q}$ holds for the Yukawa coupling of any quark, but the non-linear part is only relevant for the top
quark. This approximation is excellent over the whole range of considered scales, see Fig.~\ref{fig::ytrunning}.

\begin{figure}
\includegraphics[width=9cm]{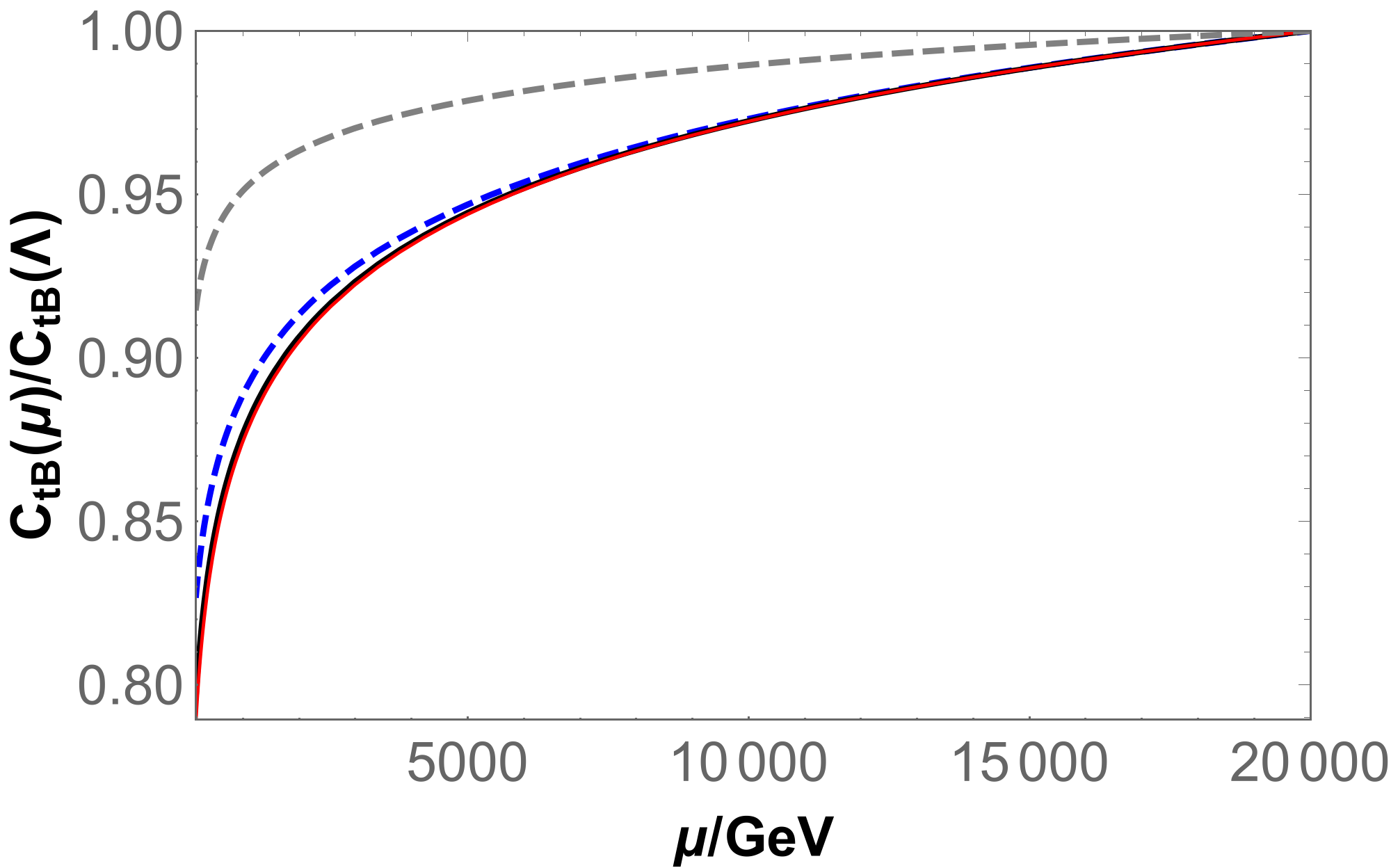}
\caption{\label{fig::CtBrunning} Running of $C_{tB}$ in different scenarios: neglecting the term $\sim y_t^2$ in the ADM (grey
dashed), assuming a scale-independent ADM (blue dashed), exact (black) and with our approximation (red, overlapping with the black
line).}
\end{figure}

Given Eq.~\eqref{eq::yqrunning}, both contributions we consider for the self mixing are proportional to $\alpha_s$ to some power.
We make these dependencies explicit by writing
\be 
\gamma_i = h^i_{\alpha_s} \alpha_s + h^i_{y_t}\alpha_s^{2\epsilon_{y_t}}\,,
\ee
such that the quantities $h^i_{\alpha_s}$ and $h^i_{y_t}$ are scale-independent, and the lower index differentiates between the two possible terms
in this equation. The integration in Eq.~\eqref{RG1} can be simplified by using 
\be
d\ln\tilde\mu=-16\pi^2\frac{d g_s}{\beta_0 g^3_s}=-2\pi\frac{d\alpha_s}{\beta_0\alpha_s^2}\,;
\ee
inserting this into Eq.~\eqref{RG} and again making the $\alpha_s$-dependence explicit, we obtain
\be\label{eq::C0alphas}
\frac{d{\cal C}_i}{d\alpha_s} =
\left(p^i_{\alpha_s}\alpha_s+p^i_{y_t}\alpha_s^{2\epsilon_{y_t}}\right)\alpha_s^{-2}\,{\cal C}_i\,,
\ee
where we introduced
\be\label{eq::epsilon}
p^i_{\alpha_s} = -\frac{h^i_{\alpha_s}}{8\pi\beta_0}\,,\qquad p^i_{y_t}=  -\frac{h^i_{y_t}}{8\pi\beta_0}\,.
\ee
In this form the integration of Eq.~\eqref{eq::C0alphas} can be performed easily, yielding
\be\label{eq::solCmu}
{\cal C}_i(\mu) = 
\eta^{p^i_{\alpha_s}} 
\exp\left[X_i(\mu)-X_i(\Lambda)\right]
{\cal C}_i(\Lambda)\,,
\ee
where 
\be
 X_i(\mu) =
 \frac{ p^i_{y_t}}{2\epsilon_{y_t}-1}[\alpha_s(\mu)]^{2\epsilon_{y_t}-1}\,,\qquad
\eta=\eta(\mu,\Lambda)\,,
\ee
with $\epsilon_{y_t}$ defined in Eq.~\eqref{eq::basicrunning2}.
This expression holds for arbitrary values of $\epsilon_{y_t}\neq1/2$. In the limit $|2\epsilon_{y_t}-1|\ll 1$, we can write
\be\label{eq::gamma0}
\gamma_i=\left(h^i_{\alpha_s}+ h^i_{y_t}\alpha_s^{2\epsilon_{y_t}-1}\right)\alpha_s\equiv h^i_{\rm eff}\,\alpha_s
\ee
and again neglect the scale dependence in the bracket so that $h^i_{\rm eff}$ is 
$\mu$-independent. In that case Eq.~\eqref{eq::solCmu} simplifies to 

\be
\label{eq::solCmuapprox}
{\cal C}_i(\mu) =
\eta^{p^i_{\rm eff}}\,
{\cal C}_i(\Lambda)\,, \qquad {p^i_{\rm eff}}=-\frac{h^i_{\rm eff}}{8\pi\beta_0}\,.
\ee
As an example, we consider top dipole operator ${\cal O}_{tB}=({\cal O}_{uB})_{33}$:
using the results from Refs.~\cite{Jenkins:2013wua,Alonso:2013hga} and Eq.~\eqref{eq::yqrunning}, we obtain
\be
h_{\alpha_s}^i = \frac{32\pi}{3}\quad {\rm and}\quad h_{y_t}^i =
\frac{15}{2}y_t(\Lambda)^2\alpha_s(\Lambda)^{-2\epsilon_{y_t}}\,.
\ee
The resulting running is shown in Fig.~\ref{fig::CtBrunning}, where we compare the exact solution Eq.~\eqref{eq::solCmu} with our
approximation in Eq.~\eqref{eq::solCmuapprox}, the solution for a constant ADM and the solution for the case where the
term $\sim y_t^2$ in the ADM is neglected: while the running due to the top-Yukawa coupling is important to include,
we see that in this case the change due to the scale dependence of the ADM is in this case actually negligible. This is due to the fact
that the scale dependence of the ADM is in this case a second-order correction: since the coefficient is present at the high scale,
already the running due to the scale-independent ADM is a correction; the relative correction to the effect we are considering is,
however, not small, it amounts to $\sim 15\%$ in this case over the considered range of scales.

\subsection{General case of one step running}
With the solutions in Eqs.~\eqref{eq::solCmu} and~\eqref{eq::solCmuapprox} at hand, we return to the RGEs for the operators induced by
the ones at the scale $\Lambda$ via weak mixing, with the goal to generalize the expressions in Section~\ref{sec:3} by including the 
scale dependence in the anomalous dimensions.

We first consider the case of a single parent operator  $\mathcal O_i$ and a single child operator $\mathcal O_k$, neglecting the
impact of mixing with other operators; the latter effect will be included below. The one-loop RGE is given as follows:
\be\label{eq::1loopRGE}
\dot{\cal C}_k=h^{ki}\,\alpha_s^{a_{ki}}{\cal C}_i+\left(h_{\alpha_s}^k\alpha_s+h_{y_t}^k
\alpha_s^{2\epsilon_{y_t}}\right){\cal C}_k\,,\quad \mbox{with}\quad k\neq i\quad\mbox{and}\quad {\cal C}_k(\Lambda)=0\,,
\ee
where again the $\alpha_s$-dependence of the anomalous dimensions is made explicit via the arbitrary exponent $a_{ki}$ and
constant parameters $h^{ki}$ and $h_{\alpha_s,y_t}^k$. The first term on the right-hand side of this equation describes
parent-child mixing, whereas the second term describes the self-mixing of the child operator.

Inserting the result for ${\cal C}_i(\mu)$ from Eq.~\eqref{eq::solCmu} we take the self-mixing of the parent operator into account. The
resulting equation can be integrated exactly. We find
\be\label{eq::Ci1loopexact}
{\cal C}_k(\mu) =
p^{ki}\,[\alpha_s(\Lambda)]^{p^k_{\alpha_s}-p^i_{\alpha_s}}\exp[X_k(\mu)-X_i(\Lambda)]\,\,\eta^{p^k_{\alpha_s}}\, \left[Y(\mu)-Y(\Lambda)\right]\,\, {\cal C}_i(\Lambda)\,,
\ee
with
\begin{align}
p^{ki}=-\frac{h^{ki}}{8\pi\beta_0}\qquad{\rm and}\phantom{blind textblind textblind textblind textblind textblind textblind f}
\end{align}
\be
Y(\mu)=\frac{1}{1-2\epsilon_{y_t}}\left(\frac{p^i_{y_t}-p^k_{y_t}}{1-2\epsilon_{y_t}}\right)^{\frac{p^i_{\alpha_s}-p^k_{\alpha_s}+a_{ki}-1}{1-2\epsilon_{y_t}}}
\hspace{-3mm}\Gamma\left(\frac{1-p^i_{\alpha_s}+p_{\alpha_s}^k-a_{ki}}{1-2\epsilon_{y_t}},\frac{[\alpha_s(\mu)]^{2\epsilon_{y_t}-1}\,(p^i_{y_t}-p^k_{y_t})}{1-2\epsilon_{y_t}}\right)\,,
\ee
where $\Gamma(s,x)$ denotes the \emph{incomplete Gamma function}, with the integral representation given as
\be
\int^v_u dt\, t^{s-1} e^{-t} =\Gamma(s,v)-\Gamma(s,u)\,.
\ee
All other symbols have been defined previously. 

\begin{figure}
\includegraphics[width=9cm]{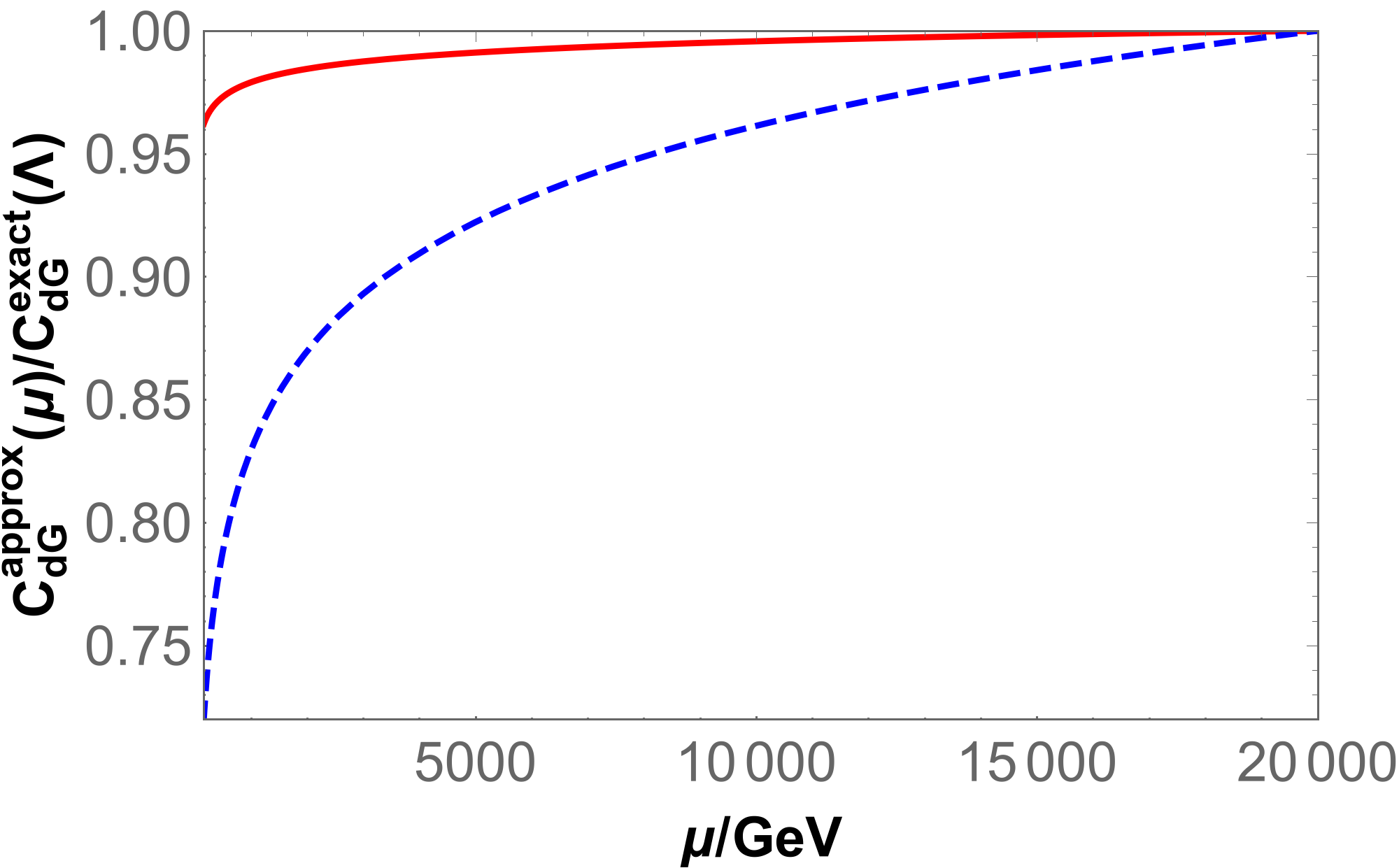}
\caption{\label{fig::CdGrunning} Comparison of the approximations discussed in the text, taking the scale-dependence of the ADM into
account (red) or not (blue dashed). Shown is the mixing of $C_{tG}$ into $C_{dG}$ relative to the exact solution.}
\end{figure}

Inserting instead the approximation Eq.~\eqref{eq::solCmuapprox},
we obtain the much simpler expression
\be\label{eq::Ci1loopapprox}
{\cal C}_k(\mu)
=\frac{p^{ki}}{a_{ki}+p_{\rm eff}^i-p_{\rm eff}^k-1}
[\alpha_s(\Lambda)]^{a_{ki}-1}\left(\eta^{a_{ki}+p_{\rm eff}^i-1}-\eta^{p_{\rm eff}^k}\right)
{\cal C}_i(\Lambda)\,,
\ee
which again is an excellent approximation. We demonstrate the different approximations in Fig.~\ref{fig::CdGrunning}, where we
consider the one-loop mixing of ${\cal O}_{tG}$ into ${\cal O}_{dG}$ as an example. As already discussed in the context of self mixing,
the influence of the scale dependence is much larger here, since there is no trivial leading-order contribution that is equal in both
cases; while the approximation of a scale-independent ADM differs from the exact solution by $\sim 30\%$ over the considered
range, our approximate formula shows less than $4\%$ deviation.

\subsection{General case of two step running}

We begin with two RG expressions
\begin{eqnarray}
\dot {\cal C}_k(\mu)&=&h^{km}\,\alpha_s^{a_{km}}{\cal C}_m(\mu)
+\left(h_{\alpha_s}^k\alpha_s+h_{y_t}^k \alpha_s^{2\epsilon_{y_t}}\right){\cal C}_k(\mu)\,,\\
\dot {\cal C}_m(\mu)&=&h^{mi}\,\alpha_s^{a_{mi}}{\cal C}_i(\mu)\,
+\left(h_{\alpha_s}^m\alpha_s+h_{y_t}^m \alpha_s^{2\epsilon_{y_t}}\right){\cal C}_m(\mu)\,, \label{eq::CmRGE}
\end{eqnarray}
together with the solution for ${\cal C}_i$ obtained previously and with boundary conditions
\be
{\cal C}_k(\Lambda)={\cal C}_m(\Lambda)=0, \qquad h^{ki}=0\,.
\ee
Evidently  $\mathcal O_k$,  $\mathcal O_m$,  $\mathcal O_i$ are child, mediator 
and parent operators, respectively. 
We solved Eq.~\eqref{eq::CmRGE} in the last subsection (for $m\to k$), so we can use for ${\cal C}_m(\mu)$ either the
exact expression in Eq.~\eqref{eq::Ci1loopexact} or the approximate one in Eq.~\eqref{eq::Ci1loopapprox}.
While the resulting equation still has an exact analytic solution within our setup in the former case, we restrict ourselves for
simplicity to the solution using the approximation in  Eq.~\eqref{eq::Ci1loopapprox}. We hence consider the equation
\be
\frac {d{\cal C}_k(\alpha_s)}{d\alpha_s}=p^{km}\,\alpha_s^{a_{km}-2}\,{\cal C}_m(\alpha_s) + p_{\rm eff}^k\,\alpha_s^{-1}\,{\cal
C}_k(\alpha_s)\,,
\ee
with ${\cal C}_m(\alpha_s)$ given in Eq.~\eqref{eq::Ci1loopapprox}. The solution reads
\be\label{eq::solRGE2loop}
{\cal C}_k(\mu) = \frac{p^{km}\,p^{mi}\,[\alpha_s(\Lambda)]^{a_{km}+a_{mi}-2}\left[A_{mi}\,\eta^{p^k_{\rm
eff}}+A_{km}\,\eta^{a_{km}+a_{mi}+p^i_{\rm eff}-2}  -A_{kmi}\,\eta^{a_{km}+p^m_{\rm eff}-1}\right]{\cal C}_i(\Lambda)}%
{A_{km}A_{mi}A_{kmi}}\,,
\ee
where we defined for convenience
\be
A_{km} = a_{km}-p^k_{\rm eff}+p^m_{\rm eff}-1\,,\quad A_{mi} = a_{mi}+p^{i}_{\rm eff}-p^m_{\rm eff}-1\,,\quad 
A_{kmi} = a_{km}+a_{mi}+p^i_{\rm eff}-p^k_{\rm eff}-2\,.
\ee

To compare to the common assumption that ${\cal C}_i$ and the anomalous dimensions $\gamma_{km,mi}$ are scale-independent, we
observe that these assumptions  correspond to the limit 
\be
p^{i,m,k}_{\rm eff}\to 0, \qquad a_{km,mi}\to 0
\ee
and yield
\be
{\cal C}_k(\mu) =
\frac{p^{km}p^{mi}}{2}
[\alpha_s(\Lambda)]^{-2}
\left[1-\eta^{-1}\right]^2 {\cal C}_i(\Lambda)\,,
\ee
with
\be
p^{km}=-\frac{\gamma_{km}}{8\pi\beta_0}\,, \qquad p^{mi}=-\frac{\gamma_{mi}}{8\pi\beta_0}\,.
\ee
Inserting Eq.~\eqref{eq::basicrunning1} one finds
\be
{\cal C}_k(\mu) =
\frac{\gamma_{km}\gamma_{mi}}{2}
\left[\frac{1}{16\pi^2}\ln\left(\frac{\Lambda}{\mu}\right)\right]^2C_i(\Lambda)\
,\ee
which can also be obtained by integrating the RGEs directly in $\ln\mu$. This serves as a
non-trivial crosscheck of Eq.~\eqref{eq::solRGE2loop}.

\subsection{Generalization to an arbitrary number of operators}

Finally we consider the case with several operators at each level, which is rather common in practical applications. Having shown the
effectiveness of our approximation \eqref{eq::approx}, we use it in the following for simplicity. As above, we start with the
self-mixing of the parent operators ${\cal O}_i$, denoting $\boldsymbol{\cal C}_P=({\cal C}_i)$ 
and analogously for child and mediator operators:
\be\label{eq::smparents}
\dot{\boldsymbol{\cal C}}_{P} = \hat\gamma^P\,\boldsymbol{\cal C}_{P} = \alpha_s\, \hat h_{\rm eff}^P\,\boldsymbol{\cal C}_{P}\,. 
\ee
Given the scale-independence of $\hat h_{\rm eff}$ in our approximation, we can easily solve this equation by diagonalization via
\be\label{eq::diag1}
\hat h_{\rm eff}^P = \hat V_P \,\hat h_{\rm eff}^{P,D}\,\hat V^{-1}_P\,,
\ee
where $\hat h_{\rm eff}^{P,D}$ is a diagonal matrix. Defining furthermore
\be\label{eq::diag2}
\boldsymbol{\cal K}_{P} = \hat V^{-1}_P\,\boldsymbol{\cal C}_{P}\,,
\ee
we obtain the equation
\be
\frac{d\boldsymbol{\cal K}_{P}}{d\alpha_s} = \alpha_s^{-1}\,\hat p^{P,D}_{\rm eff}\, \boldsymbol{\cal K}_{P}\,,
\ee
which decouples the differential equations and has the solution
\begin{eqnarray}
\boldsymbol{\cal K}_{P}(\mu) &=& \hat U_{\cal K}^P(\mu,\Lambda)\, \boldsymbol{\cal K}_{P}(\Lambda)\quad{\rm with}\quad
\hat U_{\cal K}^P(\mu,\Lambda) = \eta^{\mathbf{p}_{\rm eff}^{P,D}}\equiv {\rm diag}\left(\eta^{p_{{\rm
eff},j}^{P,D}}\right)\,,\quad{\rm or}\\
\boldsymbol{\cal C}_{P}(\mu) &=& \hat U^P(\mu,\Lambda)\, \boldsymbol{\cal C}_{P}(\Lambda)\quad{\rm
with}\quad \hat U^P(\mu,\Lambda) = \hat V_P\,\hat U_{\cal K}^P(\mu,\Lambda)\,\hat V^{-1}_P\,, \label{eq::solCPgeneral}
\end{eqnarray}
\emph{i.e.} the solution is analogous to Eq.~\eqref{eq::solCmuapprox}, only in the rotated basis. To obtain the solution in practice,
one needs to use standard linear algebra methods to obtain $\hat V_P$ and $\hat h_{\rm eff}^{M,D}$.

To generalize the one-step running, we recall Eq.~\eqref{eq::block} and hence consider
\be
\dot{\boldsymbol{\cal C}}_{C} =  \alpha_s\, \hat h_{\rm eff}^C\,\boldsymbol{\cal C}_{C} + \hat\gamma^{P\to C}\boldsymbol{\cal C}_{P}\,,\quad{\rm
where}\quad \left(\gamma_{ki}^{P\to C}\right)=\left(\alpha_s^{a_{ki}}h_{ki}^{P\to C}\right)\,.
\ee
Diagonalizing the strong mixing matrix for the child operators analogous to Eqs.~\eqref{eq::diag1} and \eqref{eq::diag2}, and inserting
the above solution for $\mathbf{C}_P$, we have
\be
\dot{\boldsymbol{\cal K}}_C = \alpha_s\, \hat h^{C,D}_{\rm eff}\,\boldsymbol{\cal K}_C+\hat V_C^{-1}\, \hat\gamma^{P\to C}\,\hat
V_P\,U_{\cal K}^P(\mu,\Lambda)\,\boldsymbol{\cal K}_P(\Lambda)\,.
\ee
This equation is already diagonal in the coefficients ${\cal K}_k$, but involves now a lengthy sum of terms with explicit $\alpha_s$
dependence, which is, however, trivial to treat in the solution of the differential equation: making the $\alpha_s$-dependence
explicit as before, we have 
\begin{eqnarray}
\frac{d{\cal K}^k_{C}}{d\alpha_s} &=& \alpha_s^{-1}\,p_{{\rm eff},k}^{C,D}\,{\cal
K}_C^k+\sum_{i,j,l}B_{klji}\,[\alpha_s(\Lambda)]^{-p_{{\rm eff},i}^{P,D}}\,\alpha_s^{a_{lj}+p_{{\rm eff},i}^{P,D}-2}{\cal
K}_P^i(\Lambda),\qquad{\rm with}\\ 
B_{klji} &=& \left(\hat V_C^{-1}\right)_{kl}p^{P\to C}_{lj}\left(\hat V_P\right)_{ji}\,.\label{eq::Bklji}
\end{eqnarray}
This equation can again be solved explicitly:
\be\label{eq::sol1stepgeneral}
{\cal K}_C^k(\mu) = \sum_{i,j,l} \frac{B_{klji}\,[\alpha_s(\Lambda)]^{a_{lj}-1}}{a_{lj}+p^{P,D}_{{\rm eff},i}-p^{C,D}_{{\rm eff},k}-1}
\left[\eta^{a_{lj}+p^{P,D}_{{\rm eff},i}-1}-\eta^{p^{C,D}_{{\rm eff},k}}\right]{\cal K}_P^i(\Lambda)\,,
\ee
which is again clearly the generalization of Eq.~\eqref{eq::Ci1loopapprox} in the (doubly) rotated basis. The solution for the coefficients of child operators in the original basis reads then
\be\label{eq::diag2C}
\boldsymbol{\cal C}_{C}(\mu)  = \hat V_C\,\boldsymbol{\cal K}_{C}(\mu)\,.
\ee

In the limit that the strong operator mixing becomes diagonal for parent and child operators, this equation simplifies to 
\be\label{simplified1}
{\cal C}_C^k(\mu) = \sum_{i} \frac{p^{P\to C}_{ki}\,[\alpha_s(\Lambda)]^{a_{ki}-1}}{a_{ki}+p^{P}_{{\rm eff},i}-p^{C}_{{\rm eff},k}-1}
\left[\eta^{a_{ki}+p^{P}_{{\rm eff},i}-1}-\eta^{p^{C}_{{\rm eff},k}}\right]{\cal C}_P^i(\Lambda)\,.
\ee
All other effects remain included, and the advantage is that the relevant coefficients can directly be read off the RGEs, without
determining the diagonalization matrices. If additionally even the diagonal running of parent or child operators can be neglected, the
corresponding $p_{\rm eff}^{(P/C),D}$ elements vanish. 

Finally, we consider two-step mixing with an arbitrary number of operators at each level. This problem is still exactly solvable,
but the equations become quite cumbersome, so we defer them to Appendix~\ref{sec::gen2step}. However, its solution
Eq.~\eqref{eq:final} provides an analytic master formula that with the help of coefficient tables obtainable from
Refs.~\cite{Jenkins:2013zja,Jenkins:2013wua,Alonso:2013hga} allows for the calculation of the two-step evolution down to the
electroweak scale, including the relvant running of all anomalous dimensions involved. Furthermore, in all these equations large
logarithms are resummed; while this might not be necessary for the weak mixing contributions, it can constitute a large effect for the
running due to the top-Yukawa and strong couplings when going to large scales $\Lambda$.
For the case of negligible self-mixing the formulae are much simpler:  we still consider
an arbitrary number of parent operators ${\mathcal O_i}$ that generate  an
arbitrary number of child operators ${\mathcal O_k}$  in a two-step running process via an arbitrary number of mediator operators
${\mathcal O_m}$, under the assumption that  there is no mixing among the operators in each group.
Then, simply using Eq.~\eqref{eq::solRGE2loop},
we find 
\be\label{eq::solRGE2loopgeneral}
{\cal C}_k(\mu) =\sum_{i,m} \frac{p^{km}p^{mi}}{a_{mi}-1}[\alpha_s(\Lambda)]^{a_{km}+a_{mi}-2}
\left[\frac{\eta^{a_{km}+a_{mi}-2}-1}{a_{km}+a_{mi}-2}-\frac{\eta^{a_{km}-1}-1}{a_{km}-1}\right] {\cal C}_i(\Lambda)\,.
\ee
While this equation still looks a bit cumbersome, it has again the advantage that all relevant coefficients can be read off
easily, without the need of diagonalization. This is also true for the case of diagonal self-mixing, discussed in
Appendix~\ref{sec::gen2step}.
Yet, one should be aware of the fact that these simpler analytic results do not describe the general case considered in
Appendix~\ref{sec::gen2step} in which all parent, mediator and child operators mix with each other. Moreover, generally the generated
operators will mix back into their respective parent operators. While the latter effect is suppressed, as discussed above, it is
interesting to note that it could in fact be included analytically as well.

\subsection{Special cases of weak mixing}\label{weakmixing}

So far we assumed an arbitrary coefficient for the $\alpha_s$-dependence of the weak ADM, parametrized by the $a^{xy}$. Here we discuss
a few specific cases in which our formalism yields even simpler expressions. 

The trivial case $a_{ki}\to 0$ is relevant for pure electroweak mixing $\sim g_{1,2}^2$ and occurs also when considering the limit
of constant ADMs. For the latter limit, we have $p_{{\rm eff},x}^X\to 0$ and $a_{ki}\to 0$ implying $p^{M\to C}_{ki}\to
-\gamma^{M\to C}_{ki}/(8\pi\beta_0)$, and we obtain from Eq.~\eqref{eq::sol1stepgeneral} the trivial generalization of
Eq.~\eqref{simplest}.

A less trivial limit is the weak mixing via Yukawa interactions. In this case, the mixing is typically proportional to $y_{q}y_{q'}$,
and its scale-dependence can be approximated by considering $a_{xy}~=~1$.
Then the general solution for one-step mixing in Eq.~(\ref{eq::sol1stepgeneral}) reduces to
\be\label{eq::sol1stepgeneralsimple}
{\cal K}_C^k = \sum_{i,j,l} \frac{B_{klji}}{p^{P,D}_{{\rm eff},i}-p^{C,D}_{{\rm eff},k}}
\left[\eta^{p^{P,D}_{{\rm eff},i}}-\eta^{p^{C,D}_{{\rm eff},k}}\right]{\cal K}_P^i(\Lambda)\,,
\ee
with $B_{klji}$ given in Eq.~(\ref{eq::Bklji}).
Similarly, the general expression for two-step mixing, given in Eq.~\eqref{eq:final}, is reduced to
\be\label{simplified4}
{\cal K}_C^k = \sum_{i,j,l,m,n,o}\frac{E_{klom}B_{mnji}\left[\Delta p_{im}\,\eta^{p_{{\rm
eff},k}^{C,D}}+\Delta p_{km}\,\eta^{p_{{\rm
eff},i}^{P,D}}-\Delta p_{ik}\,\eta^{p_{{\rm eff},m}^{M,D}}\right]{\cal K}_i^P(\Lambda)}{\Delta p_{im}\Delta p_{mk}\Delta p_{ik}}\,.
\ee
where $E_{klom}$ is defined in Eq.~\eqref{Eklom} and where we defined
\be
\Delta p_{ab} = p_{{\rm eff},a}^{X_a,D}-p_{{\rm eff},b}^{X_b,D}\,, \quad{\rm with}\quad X_a=\{P,M,C\}\,\,{\rm for}
\,\,a\in\{(i,j),(m,n,o),(k,l)\}\,.
\ee

\section{Numerical Analysis}\label{sec:5}

\begin{figure}
\includegraphics[width=8.5cm]{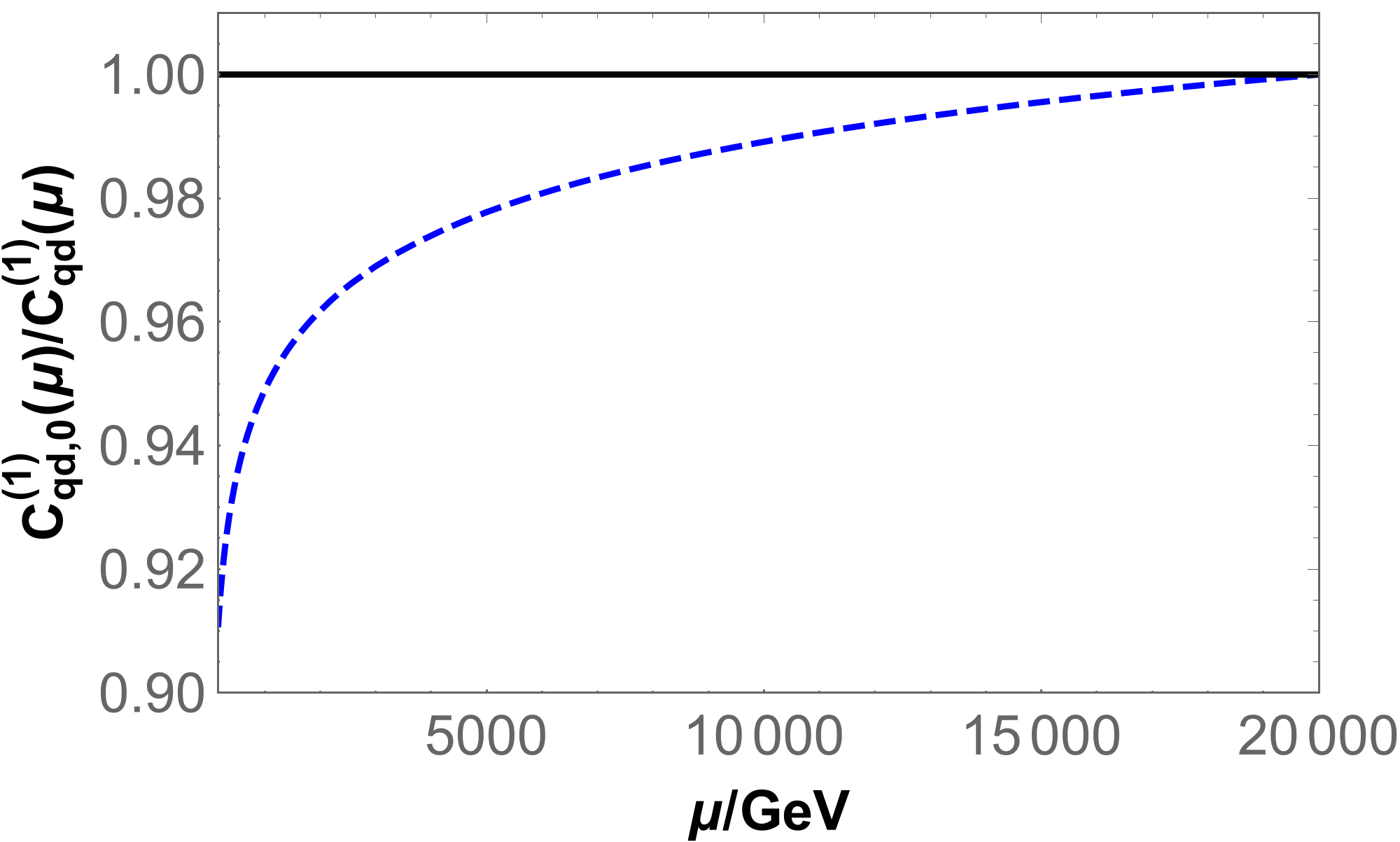}\includegraphics[width=8.5cm]{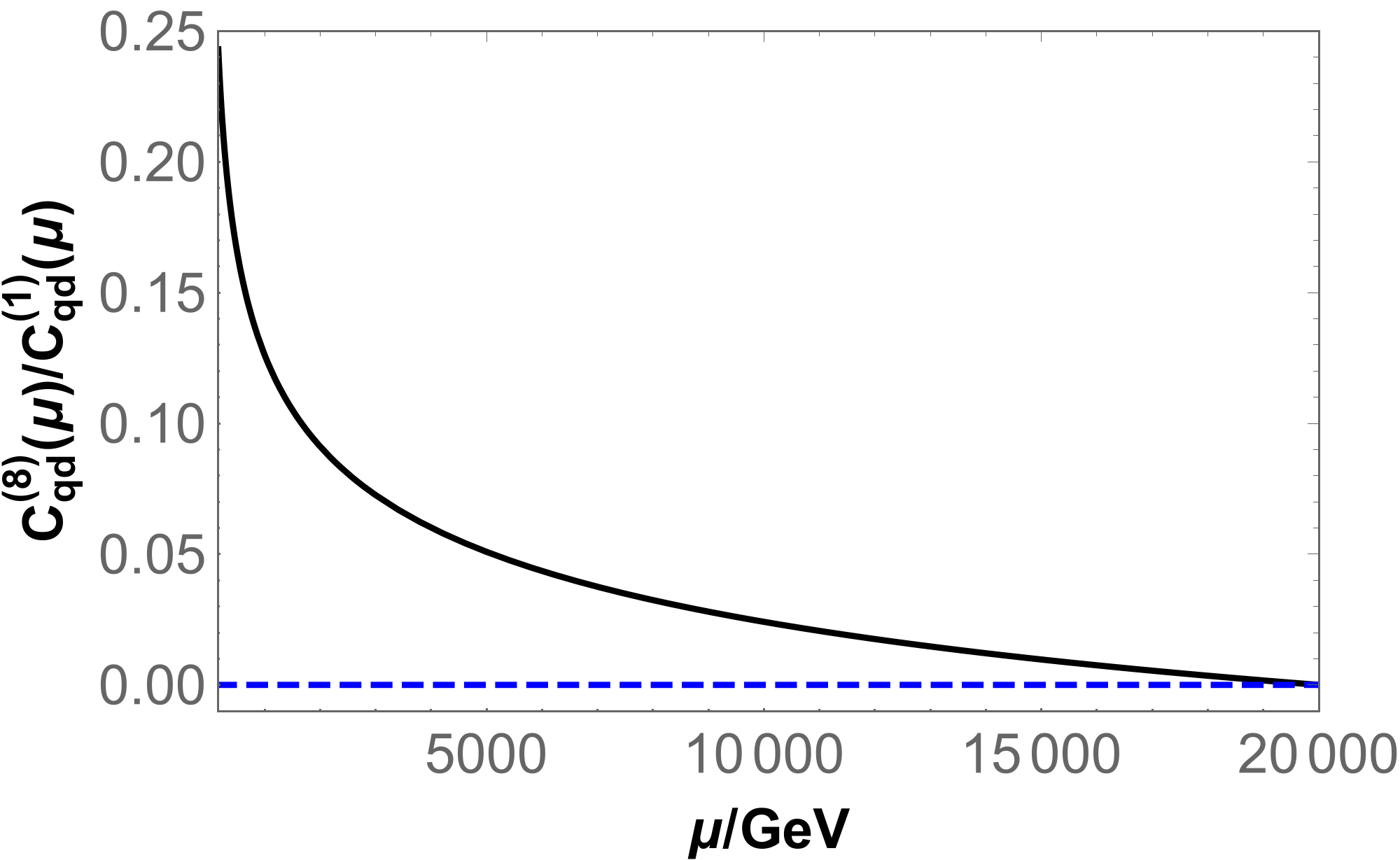}
\caption{\label{fig::example} Example of full mixing from ${\cal O}_{Hd}$ into ${\cal O}_{qd}^{(1,8)}$ beween 20~TeV and $\mu_{EW}$ in
our framework, including mixing among parent and child operators, relative to the values generated for $\hat\gamma_{P\to C}\equiv{\rm
const.},\hat\gamma_P=\hat\gamma_C=0$ (denoted by $C_{qd,0}$ and shown blue dashed) considered in \cite{Bobeth:2017xry}. }
\end{figure}

In order to facilitate the application of the formalism developed over the last sections, we perform in this section one calculation
explicitly. We choose an example where RGE running can be very important, due to the enhancement of certain (left-right) matrix
elements for meson mixing~\cite{Bobeth:2017xry}. To be even more specific, we consider the case of left-right contributions to
$B_d$-mixing, and use a basis in which the down-Yukawa matrix is diagonal, implying $Y_u=V^\dagger Y_u^{\rm diag}$ where $V$
denotes the CKM matrix.
In this specific case, only one operator ${\cal O}_{Hd}$ is created at the NP scale $\Lambda$.
Looking up its RGE \cite{Jenkins:2013wua,Alonso:2013hga}, its self-mixing is approximated via
\be
[\dot{\cal C}_{Hd}]_{13} = 6 y_t^2 [{\cal C}_{Hd}]_{13}\,,
\ee
from which we can read off, using Eqs.~\eqref{eq::smparents},\eqref{eq::diag1} and~\eqref{eq::yqrunning}, $\hat V_P =1$ and hence
\be
\hat h_{\rm eff}^{P,D}=\hat h_{\rm eff}^{P}=6 y_t(\mu_0)^2
\frac{[\alpha_s(\Lambda)]^{2\epsilon_{y_t}-1}}{[\alpha_s(\mu_0)]^{2\epsilon_{y_t}}}\,,\quad\mbox{and}\quad 
\hat p_{{\rm eff}}^{P,D}=\hat p_{{\rm eff}}^{P}=-\frac{3}{4\pi\beta_0} y_t(\mu_0)^2
\frac{[\alpha_s(\Lambda)]^{2\epsilon_{y_t}-1}}{[\alpha_s(\mu_0)]^{2\epsilon_{y_t}}}.
\ee
The child operators consist of ${\cal O}_{qd}^{(1,8)}$; the relevant parts of their RGEs read
\begin{eqnarray}
\left[\dot{\cal C}_{qd}^{(1)}\right]_{1313} &=& \left(Y_uY_u^\dagger\right)_{13} [{\cal
C}_{Hd}]_{13}+\frac{1}{2}\left(Y_uY_u^\dagger\right)_{33}\left[{\cal C}_{qd}^{(1)}\right]_{1313}-\frac{32\pi}{3}\alpha_s\left[{\cal
C}_{qd}^{(8)}\right]_{1313}\,,\\
\left[\dot{\cal C}_{qd}^{(8)}\right]_{1313} &=& \left(\frac{1}{2}\left(Y_uY_u^\dagger\right)_{33}-56\pi \alpha_s\right)\left[{\cal
C}_{qd}^{(8)}\right]_{1313}-48\pi \alpha_s\left[{\cal C}_{qd}^{(1)}\right]_{1313}\,.
\end{eqnarray}
From these equations we identify $a_{qd(1)\,Hd}=2\epsilon_{y_t}$ together with
\begin{eqnarray}
\hat p_{\rm eff}^C &=& -\frac{1}{48\pi\beta_0}\left(%
\begin{array}{c c}
3|V_{tb}|^2y_t^2/\alpha_s  &  -64\pi\\
-288\pi  &  3|V_{tb}|^2y_t^2/\alpha_s-336\pi
\end{array}\right)\equiv \left(\begin{array}{c c}x_{11} & x_{12}\\x_{21} & x_{22}

\end{array}\right)\quad{\rm and}\\
\hat p^{P\to C} &=& -\frac{1}{8\pi\beta_0}
\left(
\begin{array}{c}
V_{td}^*V_{tb}[y_t(\mu_0)]^2[\alpha_s(\mu_0)]^{-2\epsilon_{y_t}}\\0
\end{array}\right)\,.
\end{eqnarray}
The remaining diagonalization will rarely be done by hand, but for the purpose of this example we give the diagonalization matrix
and eigenvalues explicitly: defining
\be
X = \sqrt{(x_{11}-x_{22})^2+4x_{12}x_{21}}\,,
\ee
we obtain for the eigenvalues
\be
p^{C,D}_{{\rm eff},1} = \frac{1}{2}(x_{11}+x_{22}-X)\quad{\rm and}\quad p^{C,D}_{{\rm eff},2} = \frac{1}{2}(x_{11}+x_{22}+X)\,,
\ee
and for the diagonalization matrix
\be
\hat V_C = \frac{1}{2x_{21}}
\left(\begin{array}{c@{\,\,\,\,\,}c}
x_{11}-x_{22}-X & x_{11}-x_{22}+X\\
2x_{21} & 2x_{21}
\end{array}\right)\,,\quad
\hat V_C^{-1} = \frac{1}{2X}
\left(\begin{array}{c@{\,\,\,\,\,}c}
-2x_{21} & x_{11}-x_{22}+X\\
2x_{21} & -x_{11}+x_{22}+X 
\end{array}\right)\,.
\ee
With this, all the ingredients for evaluating Eq.~\eqref{eq::sol1stepgeneral} are given. In Fig.~\ref{fig::example} we show the
relative influence of the full running (within our approximations) of parent and children operators, including the mixing among the
latter, compared to the trivial running with $\hat\gamma_{P\to C}\equiv{\rm const.},\hat\gamma_P=\hat\gamma_C=0$. We observe a smaller
effect on $C_{qd}^{(1)}$ than in the previous example (see Fig.~\ref{fig::CdGrunning}), but still the influence on
$C_{qd}^{(1)}$ is $\sim 10\%$ and $C_{qd}^{(8)}$ is generated at $\sim 25\%$ of $C_{qd}^{(1)}$ at the electroweak scale when
starting from $\Lambda=20$~TeV.

\section{Summary}\label{sec:6}

We analyzed the analytic inclusion of significant contributions to the renormalization-group evolution from  the strong and
top-Yukawa couplings to operators generated via weak mixing. This can be achieved due to a distinct hierarchy between several effects,
allowing to solve the relevant RGEs in a step-wise fashion. While the running due to both the strong and the top-Yukawa coupling
(squared) can be included independently, the approximation $y_t^2(\mu)/\alpha_s(\mu)\approx {\rm const.}$ has been shown to be
excellent and facilitates the analysis greatly. The included effects are sizable mostly for the coefficients of the generated child
operators.
The analytic inclusion of these contributions facilitates for instance the application of phenomenlogical constraints at different
scales.

Our paper contains a large number of expressions that correspond to various patterns of ADMs and to different approximations.
In order to make our paper transparent we summarize the equations corresponding to these different situations in
Table~\ref{tab::summary}. In addition to the main formulae listed there, we discuss specific additional approximations in
Eqs.~\eqref{simplified1}-\eqref{simplified4} and \eqref{simplified5} that yield further simplifications.

\begin{table}
\begin{tabular}{c|c|c|c}
$\boldsymbol{P\to P}$   & \rotatebox{90}{$\gamma$ exact} & \rotatebox{90}{$\gamma$ approx.} & \rotatebox{90}{$\gamma$ const.}\\\hline
single op.              & \eqref{eq::solCmu}             & \eqref{eq::solCmuapprox}         & \eqref{RG2},\eqref{simplest}\\\hline
multi op.               & ---                            & \eqref{eq::solCPgeneral}         & \eqref{RG2},\eqref{simplest}
\end{tabular}\qquad
\begin{tabular}{c|c|c|c}
$\boldsymbol{P\to C}$   & \rotatebox{90}{$\gamma$ exact} & \rotatebox{90}{$\gamma$ approx.} & \rotatebox{90}{$\gamma$ const.}\\\hline
single op.              & \eqref{eq::Ci1loopexact}       & \eqref{eq::Ci1loopapprox}        & \eqref{RG2},\eqref{simplest}\\\hline
multi op.               & ---                            & \eqref{eq::sol1stepgeneral}      & \eqref{RG2},\eqref{simplest}
\end{tabular}\qquad
\begin{tabular}{c|c|c}
$\boldsymbol{P\to M \to C}$   &  \rotatebox{90}{$\gamma$ approx.} & \rotatebox{90}{$\gamma$ const.}\\\hline
single op.              &  \eqref{eq::solRGE2loop}         & \eqref{RG2},\eqref{eq::C3}\\\hline
multi op.               &  \eqref{eq:final}         & \eqref{RG2}
\end{tabular}   
\caption{\label{tab::summary} Summary of equations relevant for parent-parent mixing ($P\to P$), one-loop parent-child mixing
($P\to C$) and two-loop parent-mediator-child mixing ($P\to M\to C$) with different levels of approximation.}
\end{table}

We hope that the analytic results presented here will give readers better insight into the importance of the various
renormalization-group effects than it is possible by using numerical codes, only. Furthermore, the explicit expressions should allow
for quick tests of new ideas without getting involved with the intricacies of the codes present in the literature.

\section*{Acknowledgements}
This research was supported by the DFG cluster of excellence ``Origin and Structure of the Universe''.

\appendix
\section{General two-step running with an arbitrary number of operators \label{sec::gen2step}}

We consider a system of RGEs for two-step running, including strong self-mixing and an arbitrary number of operators at every stage:
\be\label{eq::RGEsystem}
\dot{\boldsymbol{\cal C}}_{P} =  \alpha_s\, \hat h_{\rm eff}^P\,\boldsymbol{\cal C}_{P}\,,\quad
\dot{\boldsymbol{\cal C}}_{M} =  \alpha_s\, \hat h_{\rm eff}^M\,\boldsymbol{\cal C}_{M} + \hat\gamma^{P\to M}\boldsymbol{\cal
C}_{P}\,,\quad
\dot{\boldsymbol{\cal C}}_{C} = \alpha_s\, \hat h_{\rm eff}^C\,\boldsymbol{\cal C}_{C} + \hat\gamma^{M\to C}\boldsymbol{\cal
C}_{M}\,,
\ee
where the three matrices $\hat h_{\rm eff}^X$ ($X=P,M,C$) are again approximately scale-independent and defined via
\be
\hat h_{\rm eff}^X = \alpha_s^{-1}\hat\gamma_X=\hat V_X \,\hat h_{\rm eff}^{X,D}\,\hat V^{-1}_X\,.
\ee
The solutions for the self-running of $\boldsymbol{\cal C}_P$ and the full expression for the mediator operators can be obtained from
Eqs.~\eqref{eq::solCPgeneral} and~\eqref{eq::sol1stepgeneral}. We rewrite the third equation in Eq.~\eqref{eq::RGEsystem} as
\be
\dot{\boldsymbol{\cal K}}_C = \alpha_s\,h^{C,D}_{\rm eff}\,\boldsymbol{\cal K}_C+\hat V_C^{-1}\,\hat\gamma^{M\to C}\,\hat V_M
\boldsymbol{\cal K}_M\,.
\ee
Inserting the solution for $\boldsymbol{\cal K}_M$, we can further rewrite this equation as
\begin{eqnarray}
\frac{d{\cal K}_C^k}{d\alpha_s} = \alpha_s^{-1}\,p_{{\rm eff},k}^{C,D}\,{\cal
K}_C^k+\!\!\!\sum_{i,j,l,m,n,o}\!\!\! E_{klom}\frac{B_{mnji}}{A_{mnji}}&&\left\{\left[\alpha_s(\Lambda)\right]^{-p_{{\rm
eff},i}^{P,D}}\alpha_s^{a_{lo}+a_{nj}+p_{{\rm eff},i}^{P,D}-3}-\right.\\
&&\,\,\,\,\left.\left[\alpha_s(\Lambda)\right]^{-p_{{\rm eff},m}^{M,D}+a_{nj}-1}\alpha_s^{a_{lo}+p_{{\rm
eff},m}^{M,D}-2}\right\}{\cal K}_i^P(\Lambda)\,,\nonumber
\end{eqnarray}
which is solved by
\be\label{eq:final}
{\cal K}_C^k =\!\!\! \sum_{i,j,l,m,n,o}\!\!\!\frac{E_{klom}B_{mnji}[\alpha_s(\Lambda)]^{a_{lo}+a_{nj}-2}\left[A_{mnji}\,\eta^{p_{{\rm
eff},k}^{C,D}}+A_{klom}\,\eta^{a_{lo}+a_{nj}+p_{{\rm
eff},i}^{P,D}-2}-A_{klonji}\,\eta^{a_{lo}+p_{{\rm eff},m}^{M,D}-1}\right]{\cal K}_i^P(\Lambda)}{A_{mnji}A_{klom}A_{klonji}}\,.
\ee
Here we introduced the abbreviations 
\begin{align}
A_{mnji} &= a_{nj}+p_{{\rm eff},i}^{P,D}-p_{{\rm eff},m}^{M,D}-1\,,\quad && A_{klom} = a_{lo}-p_{{\rm eff},k}^{C,D}+p_{{\rm
eff},m}^{M,D}-1\,,\\
A_{klonji} &= a_{lo}+a_{nj}+p_{{\rm eff},i}^{P,D}-p_{{\rm eff},k}^{C,D}-2\,,\quad && E_{klom} = (\hat V_C^{-1})_{kl} \, p_{lo}^{M\to
C}(\hat V_M)_{om}\,,\label{Eklom}
\end{align}
and used the definition in Eq.~\eqref{eq::Bklji}.
In the limit where the strong mixing matrices are diagonal, this equation simplifies to
\be\label{simplified5}
{\cal C}_C^k = \sum_{i,m}\frac{p_{km}p_{mi}[\alpha_s(\Lambda)]^{a_{km}+a_{mi}-2}\left[A_{mmii}\,\eta^{p_{{\rm
eff},k}^{C}}+A_{kkmm}\,\eta^{a_{km}+a_{mi}+p_{{\rm
eff},i}^{P}-2}-A_{kkmmii}\,\eta^{a_{km}+p_{{\rm eff},m}^{M}-1}\right]{\cal C}_i^P(\Lambda)}{A_{mmii}A_{kkmm}A_{kkmmii}}\,.
\ee
where the coefficients can be read off directly from the RGEs in \cite{Jenkins:2013zja,Jenkins:2013wua,Alonso:2013hga}.

\bibliography{BJ}
\end{document}